\documentclass[aps,prd,twocolumn,groupedaddress]{revtex4-1}
\begin{document}

%\preprint{}

\title{Spin versus helicity in processes involving transversity}

\author{Mustapha.mekhfi}
\email[]{mekhfi@gmail.com}
%\thanks{}
\altaffiliation{Theoretical Physics Division,Cern,Switzerland}
\affiliation{}


\date{\today}

\begin{abstract}
We construct the spin formalism in order to deal in a direct and natural
way with processes involving transversity which are now of increasing
popularity. The helicity formalism which is more appropriate for collision
processes of definite helicity has been so far used also to manage
processes with transversity, but at the price of computing numerous
helicity amplitudes generally involving unnecessary kinematical variables.In
a second step we work out the correspondence between both formalisms
and retrieve in another way all results of the helicity formalism
but in simpler forms.We then compute certain processes for comparison.A
special process:the quark dipole magnetic moment is shown to be exclusively
treated within the spin formalism as it is directly related to the
transverse spin of the quark inside the baryon.
\end{abstract}
\pacs{13.88.+e , 11.80.cr, 13.40.Em}
%\keywords{}
\maketitle

\section{Motivating the spin}
\noindent \textbf{}

\noindent  The amplitude of a process involving fermions of spin one half is generally written as

\begin{equation} \label{ZEqnNum748019}
\bar{u}(k,s)...X.....u(k',s{\rm '})
\end{equation}
where $u(k,s)$ is the part of the wave function which describes the spin one half of a particle of energy-momentum $p$ and spin projection $s$ and where ellipses  indicate other Dirac spinors. The probability for a given process to occur is the squared modulus of the amplitude. It is usually expressed as a trace over spinor indices by use of standard projectors

\begin{equation} \label{ZEqnNum146372}
Tr(\bar{X}\frac{\rlap{$/$}p'+m}{2m} \frac{1+\gamma _{5} \rlap{$/$}s'}{2} \cdots X\cdots \frac{\rlap{$/$}p+m}{2m} \frac{1+\gamma _{5} \rlap{$/$}s}{2} )
\end{equation}
The trace form of the probability is compact, Lorentz invariant and, in addition offers the possibility to handle the case of several $\gamma $- matrices amplitudes (several loops) using machine facilities. Several symbolic programs are made available for such symbolic computations. In the next section we will show that amplitudes as well are re-expressed as traces similarly to probabilities, and hence benefit from the same computational facilities . There exits a work similar to the present one but uses the concept of helicity( helicity formalism HF)\cite{mus1}  but not the concept of the spin ( spin formalism SF ) . Here we construct the SF and  show it  to be equivalent to the HF  . But  why the spin? if the helicity did all as we know. The answer is that the HF deals naturally with processes where the states are helicity states. It also deal with processes with states polarized along arbitrarily directions  not necessary along the momenta, but at the price of  computing  several helicity amplitudes and using unnecessary intermediary  kinematical variables . To see this, we anticipate some formulas to be shown in later sections. Let us for instance compute the electromagnetic vertex, where the initial and final particles are  polarized along  transverse directions with respect to their momenta. The vertex amplitude is of the form

\begin{equation} \label{ZEqnNum469049}
Tr(\rho _{-ss} \gamma _{\mu } )
\end{equation}
This is a one-step computation. The other way to compute the spin vertex  is to use its  decomposition  on the helicity vertices
\begin{widetext}
\begin{equation} \label{ZEqnNum977100}
Tr(\rho _{-s,s} \gamma _{\mu } )=-\frac{i}{2} \left[s(Tr(\rho _{1,1} \gamma _{\mu } )+Tr(\rho _{-1,-1} \gamma _{\mu } ))+Tr(\rho _{1,-1} \gamma _{\mu } )+Tr(\rho _{-1,1} \gamma _{\mu } )\right]
\end{equation}
\end{widetext}
\noindent
Therefore the other alternative is to compute the four helicity amplitude ( the right hand side of(\ref{ZEqnNum977100})). So treating transverse (or general) polarizations by the HF amounts rewriting such polarizations in terms of helicity states according to(\ref{ZEqnNum977100}) which  complicates the analysis somehow. On the other hand processes with transverse polarizations are no longer "academic" processes  since our revival  of the transversity  \cite{mus2} in 1992 in which we showed the relevance of the concept of transversity in hadronic physics.

\noindent A physical example ( not the computation of the vertex above) where  the HF is inappropriately and systematically used to deal with tranversity  is the one- spin transverse asymmetries . The difference of the differential cross sections are related to helicity amplitude as\cite{mus3}

\begin{equation} \label{ZEqnNum822787}
\begin{array}{rcl} {d\sigma _{\uparrow } -d\sigma _{\downarrow } } & {=} & {\left|<\cdots \left|T\right. \left|\uparrow >\, \, \right. \right|^{2} -\left|<\cdots \left|T\right. \left|\downarrow >\, \, \right. \right|^{2} } \\ {}
& {=} &2Im\left\{<\cdots \left|T\right. \left|+><+\left|T\right. \left|\cdots >\right. \right. \right\}
\end{array}
\end{equation}
 where $\left|+>\right. $and$\left|->\right. $ stand for the helicity states $\left|\pm 1/2>\right. $ and $\left|\uparrow >\right. $ and $\left|\downarrow >\right. $ for the states polarized along the transverse direction $\pm \hat{x}$. Needless to say that it is simpler to compute the first line in \ref{ZEqnNum822787} which implies computing the spin amplitude $<\cdots \left|T\right. \left|\uparrow or\downarrow >\, \, \right. $ if we can ( That is the aim of the SF)than the second line which implies computing the product of two helicity amplitudes$<\cdots \left|T\right. \left|+><+\left|T\right. \left|\cdots >\right. \right. $ a much more complex spin structure.
{}

\noindent It is the aim of the present study to give the full expression for $\rho _{ss'} (k,k')$ which is the cornerstone  of the SF which is constructed here to handle amplitudes of general polarizations . A helicity approach to a process with general polarizations not only is inappropriate but is still more cumbersome when more particles in the process are spinning, as this enhances the number of helicity components to be computed: the scattering amplitude from the generally polarized state of the electron-positron  system to a generally polarized final state of say two photons will involve sixteen helicity amplitudes $T_{\lambda }^{h} {}_{\bar{\lambda }}^{\bar{h}} $ with $h,\bar{h}$ the helicity of the photons. Even if some symmetries are present such as chirality or parity (both often absent in supersymmetic models for instance) it will not reduce the number of helicity amplitudes notably. Another issue which necessitates, but this time, an exclusive  treatment within the SF (as it depends directly on the transverse spin)is the quark dipole magnetic moment which can ultimately be written as (see  subsection 4-2)

\begin{equation} \label{6)}
\sim \int |\vec{k}|^{2} \vec{s}_{\bot } (k)\frac{d^{3} k}{(2\pi )^{3} }
\end{equation}
The above  considerations all together show definitely that the prejudice against the use of spin ( usually considered as  unessential concept  as compared to the helicity)  has no raison d'etre,  and this is sufficient  to motivate us to construct the SF

\noindent \textbf{}

\section{The amplitude as a trace}

\noindent \textbf{}

\noindent We propose in this paper to construct the SF where the spin amplitude is re-expressed as a  trace over spinor indices, as in(\ref{ZEqnNum146372}). In this way the  amplitude gets a compact form in terms of ``generalized'' projectors, thus becoming suitable for analysis and ready for symbolic computations. To this end we rewrite (\ref{ZEqnNum748019}) as (summation over repeated indices is understood).

\noindent

\begin{equation} \label{ZEqnNum938609}
\begin{array}{l} {\bar{u}_{\alpha } (k,s)\cdots f_{\alpha \alpha '} \cdots u_{\alpha '} (k',s{\rm '})} \\ {=\cdots f_{\alpha \alpha '} \cdots u_{\alpha '} (k',s{\rm '})\bar{u}_{\alpha } (k,s)} \\ {=Tr(\cdots f\rho )\, } \\ {\, \rho _{\alpha '\alpha } (k',s{\rm '},k,s)=u_{\alpha '} (k',s{\rm '})\bar{u}_{\alpha } (k,s)} \end{array}
\end{equation}
The trace involves the 4 by 4 generalized spin density matrix $\rho _{\alpha '\alpha } (k',s{\rm '},k,s)$ which when $k'=k\, \, {\rm and\; s}{\rm '}=s$ reduces to the standard projector

\begin{equation} \label{ZEqnNum795321}
(\frac{\rlap{$/$}k+m}{2m} )(\frac{1+s\gamma ^{5} \rlap{$/$}s}{2} \, )
\end{equation}
The approach we follow to work out the expression for the projector $\rho $ is to extract it from the primitive form it has at the rest frame ( relatively easy to compute)by performing  a Lorentz boost. The details of the analysis are shown in the appendix where Dirac matrices are in the standard representation and where $\gamma ^{5} =\left(\begin{array}{cc} {0} & {1} \\ {1} & {0} \end{array}\right)$. The form of the generalized density is (from now on we hide Dirac indices)
\begin{widetext}
\begin{equation} \label{ZEqnNum276244}
\rho (k',s',k,s)=\frac{2{\rm {\mathcal N}}_{s's} }{1+ss{\rm '}\vec{\zeta }.\vec{\zeta }'} (\frac{\rlap{$/$}k'+m'}{2m'} )(\frac{1+s'\gamma ^{5} \rlap{$/$}s'}{2} \, )\Re (\vec{k}',\vec{k})(\frac{\rlap{$/$}k+m}{2m} )(\frac{1+s\gamma ^{5} \rlap{$/$}s}{2} \, )
\end{equation}
\end{widetext}

\noindent The operator $\Re (\vec{k}',\vec{k})$ flips the momentum from $\vec{k}'$ to $\vec{k}$ while ${\rm {\mathcal N}}_{s's} $ a matrix in the two dimensional space of solutions ($s=\pm 1$is twice the spin)   is responsible for the spin  flip making the passage from the state $u_{\alpha } (k',s')$to the state $u_{\alpha } (k,s)$computed in the rest frame The matrix $\Re $ will be shown to have the explicit form ( see Appendix )
\begin{widetext}
\begin{equation} \label{a}
\begin{array}{rcl} {\Re (\vec{k}',\vec{k})} & {=} & {\exp (-\frac{\omega '}{2} \gamma _{0} \frac{\vec{\gamma }.\vec{k}'}{|\vec{k}'|} )\exp (\frac{\omega }{2} \gamma _{0} \frac{\vec{\gamma }.\vec{k}}{|\vec{k}|} )} \\ {} & {=} & {\cosh (\frac{\omega '}{2} )\cosh (\frac{\omega }{2} )} \\ {} & {} & {-\sinh (\frac{\omega '}{2} )\cosh (\frac{\omega }{2} )\gamma _{0} \frac{\vec{\gamma }.\vec{k}'}{|\vec{k}'|} +\sinh (\frac{\omega }{2} )\cosh (\frac{\omega '}{2} )\gamma _{0} \frac{\vec{\gamma }.\vec{k}}{|\vec{k}|} } \\ {} & {} & {+\sin h(\frac{\omega '}{2} )\sin h(\frac{\omega }{2} )\frac{\vec{\gamma }.\vec{k}'}{|\vec{k}'|} \frac{\vec{\gamma }.\vec{k}}{|\vec{k}|} } \end{array}
\end{equation}
\end{widetext}

with $\omega =-\tanh ^{-1} (\frac{|\vec{k}|}{k^{0} } )$ and idem for $\omega '$. At this point, further simplifications are possible. Since $\Re $ is inserted between energy projectors the following replacements are allowed.

\begin{equation} \label{11)}
\begin{array}{l} {\gamma _{0} \vec{\gamma }.\vec{k}=-\vec{\gamma }.\vec{k}\gamma _{0} =(\rlap{$/$}k-k_{0} \gamma _{0} )\gamma _{0} =(m\gamma _{0} -k_{0} )} \\ {\gamma _{0} \vec{\gamma }.\vec{k}'=\gamma _{0} (-\rlap{$/$}k'+\gamma _{0} k_{0} )=(k_{0} -m'\gamma _{0} )} \end{array}
\end{equation}
and idem for the primed variables. The matrix  gets then simplified and arranged in the form $\Re =A(1+\gamma _{0} )$ which becomes after some algebra

\begin{equation} \label{ZEqnNum578574}
\, \Re =\, \sqrt{\frac{mm'}{(k_{0} +m)(k'_{0} +m')} } (\gamma _{0} +1)
\end{equation}
It remains to show that formula(\ref{ZEqnNum276244}) applies equally well to massless particles i.e. it has a massless limit. To this end it is helpful to work in the center of mass frame and to adopt the normalization appropriate to massless spinors $\bar{u}(k,s)u(k,s)=2m$ which is equivalent to multiplying $\rho $ by the factor$\sqrt{2m2m'} $. We also use the fact that in the cm frame $k'_{0} =k_{0} \, ,\, \, k'=\hat{k}=(k_{0} ,-\vec{k})$ which implies that $\rlap{$/$}\hat{k}\gamma _{0} \rlap{$/$}k=\gamma _{0} \rlap{$/$}k\rlap{$/$}k=0$. With these changes appropriate to the massless case, the spin density takes the form

\begin{equation} \label{13)}
\begin{array}{l} {\sqrt{2m2m'} \rho (k',s{\rm '},k,s)|_{m'=m=0} } \\ {=\frac{1}{4k_{0} } \frac{{\rm {\mathcal N}}_{s's} }{1+s's\vec{\zeta }'.\vec{\zeta }} (1+s'\gamma ^{5} \rlap{$/$}s')\rlap{$/$}\hat{k}\, \rlap{$/$}k(1+s\gamma ^{5} \rlap{$/$}s)\, } \end{array}
\end{equation}

\noindent

\section{Spin versus helicity}

\noindent \textbf{}

\subsection{Spin versus helicity in our conventions}

\noindent

\noindent Authors of reference 1 gave explicit expressions of the helicity conserving  and non-conserving amplitudes which are the two principal components of the matrix $\rho $ computed in the HF.Their exact  forms in the center of momentum frame ($\vec{k}'=-\vec{k}$ )are (vectors $\eta$ will be defined in (\ref{ZEqnNum379105})
\begin{equation} \label{ZEqnNum379104}
\begin{array}{rcl}
\rho _{\lambda ,-\lambda } {} & {=} &u(k',\lambda )\bar{u}(k,-\lambda )=-\lambda \gamma _{0} \frac{\rlap{$/$}k+m}{2m} \frac{1-\lambda \gamma ^{5} \rlap{$/$}\eta _{3} }{2} e^{i\lambda \phi }\\
 {\rho _{\lambda ,\lambda } } & {=} & {u(k',\lambda )\bar{u}(k,\lambda )=-\lambda \gamma _{0} \frac{\rlap{$/$}k+m}{2m} \gamma _{5} \frac{\rlap{$/$}\eta ^{\lambda } }{2} e^{i\lambda \phi } } \\{}
 \rlap{$/$}\eta ^{\lambda } {} & {=} &\rlap{$/$}\eta _{1} -i\lambda \rlap{$/$}\eta _{2}
 \end{array}
\end{equation}
Here and in what follows the helicity index $\lambda $is defined with respect to the direction of $\vec{k}$, while $\lambda '$ is defined with respect to $\vec{k}'$; i.e. the frames in which  they are defined are rotated with respect to each other. The first equation in(\ref{ZEqnNum379104}) is the helicity conserving density while the second equation is the helicity non-conserving density. Note that we do not know a priori the conventions of Dirac spinors used in deriving equations(\ref{ZEqnNum379104}) and which led to the phases in the equations. But these conventions will be revealed at the end of our analysis of the SF. In this section we will extract the amplitudes (\ref{ZEqnNum379104}) from the SF and show that the conserving helicity amplitude is associated with the spin longitudinal component while the helicity flip amplitude is associated with the spin transverse component. While the first equation in(\ref{ZEqnNum379104}) is straightforward, the second equation is more subtle and need a more elaborate construction. In the HF one defines the two components spinor of definite helicity along the momentum $\vec{k}$ as ($\lambda =\pm 1$)

\begin{equation} \label{15)}
\vec{k}.\vec{\sigma }\chi _{\lambda } =\lambda \chi _{\lambda }
\end{equation}
The solution of this equation $\chi _{\lambda } $ is defined up to two independent phases each for each helicity component $\pm $ but this will be fixed promptly. To introduce the transverse spin we first introduce the ortho-normal frame ($\vec{k},\vec{\eta }_{1} ,\vec{\eta }_{2} $)

\begin{equation} \label{ZEqnNum379105}
\begin{array}{l} {\vec{k}=|\vec{k}|(\sin \theta \cos \phi ,\sin \theta \sin \phi ,\cos \theta )} \\ {\eta _{1} =(0,\vec{\eta }_{1} )=(0,\cos \theta \cos \phi ,\cos \theta \sin \phi ,-\sin \theta )} \\ {\eta _{2} =(0,\vec{\eta }_{2} )=(0,-\sin \phi ,\cos \phi ,0)} \end{array}
\end{equation}
The spinor $\chi _{1s} $ associated to the transverse spin along $\vec{\eta }_{1} $ is defined by the equation

\begin{equation} \label{17)}
\vec{\eta }_{1} .\vec{\sigma }\chi _{1s} =s\chi _{1s}
\end{equation}

\noindent

\noindent whose solution is

\begin{equation} \label{ZEqnNum689907}
\begin{array}{l} {\chi _{1}{} _{s} =\frac{(\chi _{+} +s\chi _{-} )}{\sqrt{2} } } \\ {\chi _{+} =\left(\begin{array}{c} {e^{-i\frac{\phi }{2} } \cos \frac{\theta }{2} } \\ {e^{i\frac{\phi }{2} } \sin \frac{\theta }{2} } \end{array}\right),\, \, \, \, \chi _{-} =\left(\begin{array}{c} {-e^{-i\frac{\phi }{2} } \sin \frac{\theta }{2} } \\ {e^{i\frac{\phi }{2} } \cos \frac{\theta }{2} } \end{array}\right)} \end{array}
\end{equation}
{}

\noindent

\noindent where $\chi _{+} (\chi _{-} )$ are helicity spinors along $\vec{k}$. For the need of the construction we also compute the spinor along the opposite direction $\vec{\eta }_{1} '=-\vec{\eta }_{1} $ which we denote by $\chi _{1} '{}_{s} $. To compute it, first note that $\chi _{1} '{}_{s} =\chi _{1}{} _{-s} $ then rewrite $\chi _{\lambda } $ in terms of $\chi '_{\lambda } $, that is $\chi _{-\lambda } =-i\chi '_{\lambda } $. The latter relation comes from the substitutions $\theta \to \pi -\theta ,\, \, \phi \to \phi +\pi $ made in$\chi _{\lambda } $. The spinor $\chi '_{1,s} $  thus obtained from $\chi _{1s} $ is of the form

\begin{equation} \label{ZEqnNum546453}
\chi '_{1,s} =i\frac{s\chi '_{+} -\chi '_{-} }{\sqrt{2} }
\end{equation}

\noindent The spinor $\chi _{2s} $ along $\vec{\eta }_{2} $ can be worked out similarly. We may check that it is has the form

\noindent

\begin{equation} \label{20)}
\begin{array}{l} {\chi _{2}{} _{s} =\frac{(\chi _{+} +is\chi _{-} )}{\sqrt{2} } } \\ {\chi _{2} *{}_{s} \vec{\sigma }\chi _{2} {}_{s} =s\vec{\eta }_{2} } \end{array}
\end{equation}
and finally the spinor $\chi _{2} '{}_{s} $ can be extracted from $\chi _{2} {}_{s} $ similarly to$\chi _{1} '{}_{s} $ with the result

\begin{equation} \label{21)}
\chi _{2,} '{}_{s} =i\frac{(is\chi '_{+} -\chi '_{-} )}{\sqrt{2} }
\end{equation}
These spinors being at hand we may compute the various elements of the matrix ${\rm {\mathcal N}}$ which occurs in our principal formula(\ref{ZEqnNum276244})

\noindent

\begin{equation} \label{ZEqnNum450454}
\begin{array}{l} {{\rm {\mathcal N}}_{-\lambda ,\lambda } =\chi '_{-\lambda } {}_{j} \chi ^{\dag } _{\lambda j} =-i\chi _{\lambda j} \chi ^{\dag } _{\lambda j} =-i} \\ {{\rm {\mathcal N}}_{{\rm 1},} {}_{-s,s} ={\rm {\mathcal N}}_{2,}{} _{-s,s} =-1} \end{array}
\end{equation}
We are now ready to use the SF in order to re-uncover the results (\ref{ZEqnNum379104}). The use of the SF, among others,  may be considered as another way of re-deriving the helicity formulas, first derived in reference 1, from a generalization of a formula due to the Michel and Bouchiat\cite{mus4} .Recall that the formulas (\ref{ZEqnNum379104}) are computed in the centre of mass frame, so computations in this section will be performed in the centre of mass frame.  Let us first compute the helicity conserving case $(\lambda '=\lambda )$using our formula(\ref{ZEqnNum276244}). To this end we make the assignments:

\begin{equation} \label{ZEqnNum667934}
\begin{array}{l} {\vec{\zeta }'=-\vec{\zeta }} \\ {\vec{\zeta }=\frac{\vec{k}}{{\rm | }\vec{k}{\rm | }} } \end{array}
\end{equation}
Within these specifications the spin indices $s{\rm ',}s$ become helicity indices respectively $\lambda {\rm ',}\lambda $ .

\begin{widetext}
\begin{equation} \label{ZEqnNum738865}
\begin{array}{rcl} {} & {} & {u(k',\lambda ')\bar{u}(k,\lambda )} \\ {} & {=} & {(\frac{\hat{\rlap{$/$}k}+m}{2m} )(\frac{1+\lambda '\gamma ^{5} \hat{\rlap{$/$}s}_{L} }{2} \, ){\rm {\mathcal N}}_{\lambda ',\lambda } (\vec{\zeta }'=-\vec{\zeta },\vec{\zeta }\, )\Re (\hat{k},k)(\frac{\rlap{$/$}k+m}{2m} )(\frac{1+\lambda \gamma ^{5} \rlap{$/$}s_{L} }{2} \, )} \end{array}
\end{equation}
\end{widetext}

The various factors in(\ref{ZEqnNum738865}) computed in the centre of mass frame take the forms

\begin{equation} \label{ZEqnNum427055}
\begin{array}{l} {\hat{\rlap{$/$}s}_{L} =\frac{|\vec{k}|}{m} \gamma _{0} +\frac{k_{0} }{m} \frac{\vec{k}.\vec{\gamma }}{|\vec{k}|} } \\ {{\rm \Re (}\vec{k}'=-\vec{k},\vec{k})=\frac{m}{k_{0} +m} (1+\gamma _{0} )} \\ {{\rm {\mathcal N}}_{\lambda ,-\lambda } =-i} \end{array}
\end{equation}

\noindent Inserting (\ref{ZEqnNum427055}) into (\ref{ZEqnNum738865}) and moving the projector to the left of the factor$(1+\gamma _{0} )$ with the help of the identity ($\hat{k}=(k_{0} ,-\vec{k})$)

\begin{equation} \label{26)}
\frac{\hat{\rlap{$/$}k}+m}{2m} (1+\gamma _{0} )\frac{\rlap{$/$}k+m}{2m} =\frac{m+k_{0} }{m} \gamma _{0} \frac{\rlap{$/$}k+m}{2m}
\end{equation}
 We get the final expression

\begin{equation} \label{ZEqnNum842342}
\begin{array}{rcl} {u(k',\lambda )\bar{u}(k,-\lambda )}  = {-i\gamma _{0} (\frac{\rlap{$/$}k+m}{2m} )(\frac{1-\lambda \gamma ^{5} \rlap{$/$}s_{L} }{2} \, )} \end{array}
\end{equation}
To retrieve the second equation in (\ref{ZEqnNum379104}) much more work is needed. We first adopt the new assignments in which the polarizations $\vec{\zeta }',\vec{\zeta }$ are oriented perpendicularly to $\vec{k}$

\begin{equation} \label{ZEqnNum632421}
\begin{array}{l} {\vec{\zeta }'=-\vec{\zeta }} \\ {\vec{\zeta }'.\vec{k}=\vec{\zeta }.\vec{k}=0} \end{array}
\end{equation}
In the sequel we will need set $\vec{\zeta }=\vec{\eta }_{1} $ and then $\vec{\zeta }=\vec{\eta }_{2} $.The transverse component of the generalized spin density is
\begin{widetext}
\begin{equation} \label{ZEqnNum547620}
\begin{array}{rcl} {} & {} &u(k',-s)\bar{u}(k,s)|_{\vec{\zeta }}\\
 & {=} & (\frac{\rlap{$/$}\hat{k}+m}{2m} )(\frac{1-s\gamma ^{5} \rlap{$/$}s'_{\bot } }{2} \, ){\rm {\mathcal N}}_{-s,s} (\vec{\zeta }'=-\vec{\zeta },\vec{\zeta }\, )\Re (\vec{k}',\vec{k})(\frac{\rlap{$/$}k+m}{2m} )(\frac{1+s\gamma ^{5} \rlap{$/$}s_{\bot } }{2} \, )\end{array}
\end{equation}
\end{widetext}

The various factors in (\ref{ZEqnNum547620}) computed in the centre of mass frame take the forms

\begin{equation} \label{ZEqnNum790401}
\begin{array}{l} {\vec{\zeta }\equiv \vec{\eta }_{1} ,\vec{\eta }_{2} } \\ {{\rm {\mathcal N}}_{-s,s} =-1} \\ {\rlap{$/$}s_{\bot } =-\vec{s}_{\bot } .\vec{\gamma }} \\ {\rlap{$/$}s'_{\bot } =-\rlap{$/$}s_{\bot } } \end{array}
\end{equation}
Inserting the values(\ref{ZEqnNum790401}) into(\ref{ZEqnNum547620}) and performing some algebra we get the simplified expression

\begin{equation} \label{ZEqnNum392998}
u(k',-s)\bar{u}(k,s)|_{\vec{\zeta }} =-\gamma _{0} (\frac{\rlap{$/$}k+m}{2m} )(\frac{1+s\gamma ^{5} \rlap{$/$}s_{\bot } }{2} \, )
\end{equation}

\noindent Now  define the quantity $\, \Delta (\vec{\zeta },s)$ and compute it
\vspace{7mm}
\begin{widetext}
\begin{equation} \label{ZEqnNum300809}
\begin{array}{rcl} {\, \Delta (\vec{\zeta },s)} & {=} & {u(k',-s)\bar{u}(k,s)|_{\vec{\zeta }=\vec{\eta }_{1} ,\vec{\eta }_{2} } -\frac{1}{2} \sum _{s=\pm 1}u(k',-s)\bar{u}(k,s)|_{\vec{\zeta }=\vec{\eta }_{1} ,\vec{\eta }_{2} }  } \\ {}
& {=} &-\gamma _{0} (\frac{\rlap{$/$}k+m}{2m} )\frac{s\gamma ^{5} \rlap{$/$}\zeta }{2} |_{\vec{\zeta }=\vec{\eta }_{1} ,\vec{\eta }_{2} } \,
\end{array}
\end{equation}
\end{widetext}
                                                                                                                                                                                  To link with the HF we project the transverse spin components $u(k,s)|_{\vec{\zeta }=\vec{\eta }_{1} } $ etc, on the helicity components along $\vec{k}$ and $\vec{k}'$ : $u(k,\lambda )$ and $u(k',\lambda ')$. It can be shown that these projections are similar to those of spinors at rest. The latter have been worked previously. We just replace spinors at rest by Dirac spinors. The projections are as follows

\begin{equation} \label{ZEqnNum794799}
\begin{array}{l} {u(k,s)|_{\vec{\zeta }=\vec{\eta }_{1} } =\frac{u(k,+1)+su(k,-1)}{\sqrt{2} } } \\ {u(k',-s)|_{\vec{\zeta }=\vec{\eta }_{1} }  =  -i\frac{u(k',-1){\rm +}su(k',+1)}{\sqrt{2} } } \\ {u(k,s)|_{\vec{\zeta }=\vec{\eta }_{2} }   = \frac{u(k,+1)+isu(k,-1)}{\sqrt{2} } } \\ {u(k',-s)|_{\vec{\zeta }=\vec{\eta }_{2} }  = -i\frac{u(k',-1){\rm +{\tt i}}su(k',+1)}{\sqrt{2} } } \end{array}
\end{equation}
\vspace{7mm}

Back to(\ref{ZEqnNum300809}) and using(\ref{ZEqnNum794799}) we get
\begin{widetext}
\begin{equation} \label{ZEqnNum465965}
\begin{array}{rcl} {\, i\Delta (\vec{\eta }_{1} ,s)} & {=} & {iu(k',-s)\bar{u}(k,s)|_{\vec{\zeta }=\vec{\eta }_{1} } -\frac{1}{2} i\sum _{{\rm {\tt s}}=\pm 1}u(k',-s)\bar{u}(k,s)|_{\vec{\zeta }=\vec{\eta }_{1} }  } \\ {}
& {=} &s\frac{u(k',+1)\bar{u}(k,+1)+u(k',-1)\bar{u}(k,-1)}{2} \\

i\Delta (\vec{\eta }_{2} ,s)& {=} &iu(k',-s)\bar{u}(k,s)|_{\vec{\zeta }=\vec{\eta }_{2} } -\frac{1}{2} i\sum _{{\rm {\tt s}}=\pm 1}u(k',-s)\bar{u}(k,s)|_{\vec{\zeta }=\vec{\eta }_{2} } \\

& {=} &is\frac{u(k',+1)\bar{u}(k,+1)-u(k',-1)\bar{u}(k,-1)}{2}

\end{array}
\end{equation}
\end{widetext}

 The following linear combination is the desired expression which according to (\ref{ZEqnNum465965}) and(\ref{ZEqnNum392998}) has the form

\begin{equation} \label{ZEqnNum668855}
\begin{array}{rcl} {is(\Delta (\vec{\eta }_{1} ,s)-i\lambda \Delta (\vec{\eta }_{2} ,s))} & {=} & {u(k',\lambda )\bar{u}(k,\lambda )} \\ {} & {=} & {-i\gamma _{0} (\frac{\rlap{$/$}k+m}{2m} )\gamma ^{5} \frac{\rlap{$/$}\eta ^{\lambda } }{2} \, } \end{array}
\end{equation}

\subsection { Translating our results to the conventions of reference 1}

\noindent

\noindent Equations(\ref{ZEqnNum842342}) and(\ref{ZEqnNum668855}) are the relationships between the spin projectors and the helicity ones written with our choice of spinors and our orientation of the transverse spin around $\vec{k}$. To retrieve the helicity projectors of equations(\ref{ZEqnNum379104}) we have to convert to the conventions of reference1.

\noindent Inspection of  the first equation in(\ref{ZEqnNum379104}) shows that the spinors (at rest) $\tilde{\chi }_{\lambda } $ used by the author of reference 1 are related to ours by $\tilde{\chi }_{\lambda } =\lambda e^{i\lambda \frac{\phi }{2} } \chi _{\lambda } $and this amounts to rewrite(\ref{ZEqnNum842342}) as

\begin{equation} \label{36)}
\begin{array}{rcl} {} & {} & {-u(k',\lambda )\bar{u}(k,-\lambda )} \\ {} & {=} & {\lambda e^{i\lambda \phi } \gamma _{0} (\frac{\rlap{$/$}k+m}{2m} )(\frac{1-\lambda \gamma ^{5} \rlap{$/$}s_{L} }{2} \, )} \end{array}
\end{equation}
Wherein (\ref{ZEqnNum842342}) ${\rm {\mathcal N}}_{\lambda ,-\lambda } =-i$ has been replaced by ${\rm {\mathcal N}}_{\lambda ,-\lambda } =\lambda e^{i\lambda \phi } $ and  $u(k',\lambda )\bar{u}(k,-\lambda )$ multiplied by $\lambda (-\lambda )=-1$, hence retrieving the first equation in (\ref{ZEqnNum379104}).   On the other hand inspection of  the  second equation in(\ref{ZEqnNum379104}) shows that the combination

\begin{equation} \label{37)}
e^{i\lambda \phi } \rlap{$/$}\eta ^{\lambda }
\end{equation}
is an indication that the trihedral $\vec{\eta }_{1} ,\vec{\eta }_{2} ,\vec{k}$ we used has to be rotated around $\vec{k}$ by an angle $\chi $,  an angle to be fixed,  in addition to the change of spinors. The first operation amounts to multiply $u(k',\lambda )\bar{u}(k,\lambda )$ in(\ref{ZEqnNum668855}) by the phase factor $e^{i\lambda \chi } $ and the combination $\rlap{$/$}\eta ^{\lambda } $by$e^{-i\lambda \chi } $ as the scalar product here $\eta _{\mu } ^{\lambda } .\bar{u}(k,\lambda )\gamma ^{\mu } u(k',\lambda )$ is invariant under rotations. The second operation which is the change of spinors amounts to multiplying  in(\ref{ZEqnNum668855})$u(k',\lambda )\bar{u}(k,\lambda )$ by the factor $e^{i\lambda \phi } $and  the ${\rm {\mathcal N}}_{-s,s} =-1$ factor accompanying  $\eta $  has been replaced by ${\rm {\mathcal N}}_{-s,s} =-i\lambda e^{i\lambda \phi } $.
Inserting both modifications into(\ref{ZEqnNum668855}) gives

\begin{equation} \label{ZEqnNum536153}
\begin{array}{rcl} {is(\Delta (\vec{\eta }'_{1} ,s)-i\lambda \Delta (\vec{\eta }'_{2} ,s))} & {=} & {e^{i\lambda \chi } u(k',\lambda )\bar{u}(k,\lambda )} \\ {} & {=} & {\lambda \gamma _{0} e^{-i\lambda \chi } (\frac{\rlap{$/$}k+m}{2m} )\gamma ^{5} \frac{\rlap{$/$}\eta ^{\lambda } }{2} \, } \end{array}
\end{equation}

\noindent If we fix the rotation angle to$\chi =(-\phi +\pi )/2$ then we recover the helicity projectors as they appeared in reference1, hence ending up the comparison.

\noindent

\begin{equation} \label{40)}
\begin{array}{rcl} {is(\Delta (\vec{\eta }'_{1} ,s)-i\lambda \Delta (\vec{\eta }'_{2} ,s))} & {=} & {u(k',\lambda )\bar{u}(k,\lambda )} \\ {} & {=} & {-\lambda \gamma _{0} e^{i\lambda \phi } (\frac{\rlap{$/$}k+m}{2m} )\gamma ^{5} \frac{\rlap{$/$}\eta ^{\lambda } }{2} \, } \end{array}
\end{equation}

\noindent
\subsection{principal formulas of the HF and the SF}

\noindent we recapitulate here the principal results of the SF and the HF written in the center of mass frame. The HF formulas are

\begin{equation} \label{1}
\begin{array}{rcl}
\rho _{\lambda ,-\lambda } {} & {=} &u(k',\lambda )\bar{u}(k,-\lambda )=-\lambda \gamma _{0} \frac{\rlap{$/$}k+m}{2m} \frac{1-\lambda \gamma ^{5} \rlap{$/$}\eta _{3} }{2} e^{i\lambda \phi }\\
 {\rho _{\lambda ,\lambda } } & {=} & {u(k',\lambda )\bar{u}(k,\lambda )=-\lambda \gamma _{0} \frac{\rlap{$/$}k+m}{2m} \gamma _{5} \frac{\rlap{$/$}\eta ^{\lambda } }{2} e^{i\lambda \phi } } \\{}
 \rlap{$/$}\eta ^{\lambda } {} & {=} &\rlap{$/$}\eta _{1} -i\lambda \rlap{$/$}\eta _{2}
 \end{array}
\end{equation}
 and the SF in our conventions are
 \begin{equation} \label{2}
\begin{array}{rcl}
 \rho _{s ,s }|_{\vec{k}'} {} & {=} &u(k',s)\bar{u}(k,s)|_{\vec{k }'}= {-i\gamma _{0} (\frac{\rlap{$/$}k+m}{2m} )(\frac{1-s \gamma ^{5} \rlap{$/$}s_{L} }{2} \, )} \\
\rho _{-s ,s }|_{\vec{\zeta }}{} & {=} &u(k',-s)\bar{u}(k,s)|_{\vec{\zeta }} = -\gamma _{0} (\frac{\rlap{$/$}k+m}{2m} )(\frac{1+s\gamma ^{5} \rlap{$/$}s_{\bot } }{2} \, )
\end{array}
\end{equation}
and the links between both formalism are

\begin{equation} \label{40)}
\begin{array}{rcl}
\rho _{s ,s }|_{\vec{k}'} =\rho _{\lambda ,-\lambda }\\
{is(\Delta (\vec{\eta }'_{1} ,s)-i\lambda \Delta (\vec{\eta }'_{2} ,s))} &{=} & {u(k',\lambda )\bar{u}(k,\lambda )}   \end{array}
\end{equation}

\section{Computing a given process in the SF}

\subsection{ $e^{-} \to e^{-} +\gamma $and $e^{+} e^{-} \to \tilde{e}^{+} _{R} \tilde{e}^{-} _{R} $}

\noindent Let us compute the simplest possible amplitude in the SF and in the HF as an illustration of the performance of the spin over the helicity in the presence of  a general polarization. So a vertex such as $e^{-} \to e^{-} +\gamma $or even simpler such as $e^{-} \to \tilde{e}+\tilde{\gamma }$ where $(\tilde{e},\, \tilde{\gamma })$ is a (slectron, photino) system, may serve as an example of such compuations. The decomposition of the spin state over the helicity states (\ref{ZEqnNum794799}) is equivalent to the spin projector $\rho _{-ss} $ decomposed over the helicity projectors

\begin{equation} \label{ZEqnNum621678}
\rho _{-s,s} =-\frac{i}{2} \left[s(\rho _{1,1} +\rho _{-1,-1} )+\rho _{1,-1} +\rho _{-1,1} \right]
\end{equation}
The electromagnetic vertex is simply the trace

\begin{equation} \label{42)}
Tr(\rho _{-ss} \gamma _{\mu } )
\end{equation}

\noindent If we chose to compute the vertex in the SF, this single trace is all we have to compute. The computation of the same trace but in the HF necessitates computing   all the  helicity projections of the right hand side of(\ref{ZEqnNum621678}). Acknowledge that this is a long computation. So by using the SF, that is the second equation in (\ref{2}) the computation is straightforward

\begin{equation} \label{ZEqnNum163291}
2iTr(\rho_{-s,s} \gamma _{\mu } )=-\frac{2s}{m} \left|\vec{k}\right|\eta _{1 \mu } -2ig_{\mu 0}
\end{equation}
But preferring the HF instead, one has to compute four helicity components using the helicity projectors,(\ref{1}) and then (\ref{ZEqnNum621678}) before recovering the result. The four helicity components are

\begin{equation} \label{ZEqnNum295347}
\begin{array}{l} {Tr(\rho _{\lambda ,\lambda } \gamma _{\mu } )=-\frac{i\lambda }{m} \left|\vec{k}\right|\eta ^{\lambda } _{\mu } } \\ {Tr(\rho _{-\lambda ,\lambda } \gamma _{\mu } )=-ig_{\mu 0} } \end{array}
\end{equation}
Note that the vertex amplitude(\ref{ZEqnNum163291}) involves the transverse vector $\eta _{1\mu } $ only, while the helicity components(\ref{ZEqnNum295347}) involve in addition  $\vec{\eta }_{2} $ and $\vec{s}_{L} $ $(\vec{k}\times \vec{s}_{L} =0)$  but as unessential intermediary steps in the computation as they do not appear in the final result. This clearly shows that the HF is not appropriate in the presence of transverse spin.

\noindent Let us compute a more involved process  $e^{+} e^{-} \to \tilde{e}^{+} _{R} \tilde{e}^{-} _{R} $ of the creation of right-handed selectrons by electrons via exchange of photinos in the chiral case  ( $photons$,$Z-bosons$ and $zino$  exchanges  are not considered as this computation is just an illustration).The positron has momentum $k$ along the $z$ axis in the $e^{+} e^{-} $centre of mass system  and spin projection $-s$ along the transverse direction $\eta _{1} $ (here the $ x $-axis), while the electron has momentum $k'$ and spin $-s$ along  $\eta '_{1} $ opposite to $\eta _{1} $( this is the natural  polarization of the $e^{+} e^{-} $system  in the storage ring). The momentum of the selectron is $p=(p^{0} ,\vec{p})\, $ with $\vec{p}=\left|\vec{p}\right|(\sin \theta \cos \phi ,\sin \theta \sin \phi ,\cos \theta )$. We may replace the electron-positron system by the quark-antiquark system with transverse polarization inherited from the parent proton-antiproton system transversely polarized. For the electron-photino vertex, we use the vertex fixed by super-symmetry in the chiral case $\sqrt{2\, } e\bar{\tilde{\gamma }}\frac{(1+\gamma _{5} )}{2} e\tilde{e}*_{R} $.Now to compute the amplitude we have first to include the anti- particle into the formalism  since so far we have  considered  only the projector involving particles. To do so we adopt the convention that the anti-particle spinor is related to the particle spinor by the relation\cite{mus5}

\noindent

\begin{equation} \label{ZEqnNum584301}
v(k,\lambda )=-\lambda \gamma _{5} u(k,-\lambda )
\end{equation}
Applying the relation(\ref{ZEqnNum584301}) to the spinor $u(k,s)|_{\vec{\zeta }=\vec{\eta }_{1} } $ in(\ref{ZEqnNum794799}) we get

\begin{equation} \label{46)}
\begin{array}{rcl} {u(k,s)|_{\vec{\zeta }=\vec{\eta }_{1} } } & {=} & {\frac{u(k,1)+su(k,-1)}{\sqrt{2} } } \\ {}
& {=} &s\gamma _{5} \frac{v(k,1)-sv(k,-1)}{\sqrt{2} }\\

& {=} &s\gamma _{5} v(k,-s)|_{\vec{\zeta }=\vec{\eta }_{1} }
 \end{array}
\end{equation}

 \noindent In what follows all quantities will be computed using the energy projectors $\rlap{$/$}k+m$  (suitable for the massless limit or high energy which we will adopt here).

\noindent The amplitude computed at vanishing azimuthal angle is of the form

\noindent

\begin{equation} \label{47)}
\begin{array}{l} {2e^{2} \bar{v}(k,-s)\frac{(1-\gamma _{5} )}{2} (\frac{\rlap{$/$}k'-\rlap{$/$}p+m_{\tilde{\gamma }} }{t-m^{2} _{\tilde{\gamma }} } )\frac{(1+\gamma _{5} )}{2} u(k',-s)} \\ {=\frac{se^{2} }{2(t-m^{2} _{\tilde{\gamma }} )} Tr(1-\gamma _{5} )(\rlap{$/$}k'-\rlap{$/$}p)\gamma _{0} \rlap{$/$}k(1+s\gamma ^{5} \rlap{$/$}\eta _{1} \, )} \end{array}
\end{equation}
with $t=(k'-p)^{2} $. The $\gamma _{5} $term is vanishing as it is proportional to $\vec{p}.(\vec{k}'\times \vec{\eta }_{1} )=\vec{p}.\vec{\eta }_{2} =0 $while the remaining term is

\begin{equation} \label{ZEqnNum849442}
\begin{array}{l} {\frac{e^{2} }{2(t-m^{2} _{\tilde{\gamma }} )} Tr(\rlap{$/$}k'-\rlap{$/$}p)\gamma _{0} \rlap{$/$}k\rlap{$/$}\eta _{1} \, } \\ {=\frac{2e^{2} k_{0}^{2} \beta _{R} \sin \theta }{(t-m^{2} _{\tilde{\gamma }} )} \, } \end{array}
\end{equation}

\noindent with $\beta _{R} =\frac{\left|\vec{p}\right|}{p^{0} } $ the velocity of the selectron. This is a one-step computation in the SF. To compute the same amplitude but in the HF as is commonly practiced so far, we have to compute four helicity amplitudes separately. The amplitude of the above process computed in the HF involves the quantities

\begin{equation} \label{49)}
\begin{array}{l} {A_{\mu } =\bar{u}(k',\lambda ')\gamma _{\mu } v(k,\lambda )} \\ {B_{\mu } =\bar{u}(k',\lambda ')\gamma _{\mu } \gamma _{5} v(k,\lambda )} \end{array}
\end{equation}
Computed in our convention, they lead to the expressions

\begin{equation} \label{50)}
\begin{array}{l} {A_{\mu ,\lambda ,\lambda '} =\left\{\begin{array}{c} {-i2mg_{\mu i} \hat{k}'^{i} \, \, \, \, \, \, \lambda =\lambda '} \\ {-i2k'_{0} \lambda '\eta _{\mu }^{\lambda } \, \, \, \, \, \, \, \, \, \, \, \, \lambda =-\lambda '} \end{array}\right. } \\ {B_{\mu ,\lambda ,\lambda '} =\left\{\begin{array}{c} {-i2\left|\vec{k}'\right|\eta ^{\lambda } _{\mu } \, \, \, \lambda =-\lambda '} \\ {-i2m\lambda 'g_{\mu 0\, \, \, \, \, \, \, \, \, } \, \, \, \, \, \, \, \, \, \, \, \, \lambda =\lambda '} \end{array}\right. } \end{array}
\end{equation}

\noindent The helicity amplitudes turn out to all vanish in the limit of vanishing electron mass, except the one associated to the right-handed electron which we write explicitly

\begin{equation} \label{51)}
\begin{array}{l} {2e^{2} \bar{v}(k,-1)\frac{(1-\gamma _{5} )}{2} (\frac{\rlap{$/$}k'-\rlap{$/$}p+m_{\tilde{\gamma }} }{t-m^{2} _{\tilde{\gamma }} } )\frac{(1+\gamma _{5} )}{2} u(k',1)} \\ {=\frac{e^{2} (k'-p)^{\mu } }{t-m^{2} _{\tilde{\gamma }} } (A^{*} _{\mu ,1,-1} +B^{*} _{\mu ,1,-1} )} \\ {=-ie^{2} \frac{4k_{0} p.n_{\mu }^{\lambda } }{t-m^{2} _{\tilde{\gamma }} } } \\ {=-ie^{2} \beta _{R} \frac{4k^{2} _{0} \sin \theta }{t-m^{2} _{\tilde{\gamma }} } } \end{array}
\end{equation}
 Now using the decomposition of the spin projector $\tilde{\rho }_{s,s'} =u(k,s)\bar{v}(k',s')$ along the helicity projectors analogous to(\ref{ZEqnNum621678})

\begin{equation} \label{ZEqnNum707111}
\tilde{\rho }_{-s,-s} =-\frac{i}{2} \left[s(\tilde{\rho }_{1,1} -\tilde{\rho }_{-1,-1} )+\tilde{\rho }_{-1,1} -\tilde{\rho }_{1,-1} )\right]
\end{equation}
 We recover the transversely polarized amplitude which is according to(\ref{ZEqnNum707111})

\begin{equation} \label{53)}
\begin{array}{rcl} {\frac{i}{2} Tr(\tilde{\rho }_{1,-1} \cdots )} & {=} & {\frac{i}{2} (-ie^{2} \beta _{R} \frac{4k^{2} _{0} \sin \theta }{t-m^{2} _{\tilde{\gamma }} } )} \\ {} & {=} & {\frac{2e^{2} k_{0}^{2} \beta _{R} \sin \theta }{(t-m^{2} _{\tilde{\gamma }} )} \, } \end{array}
\end{equation}
The result is recovered but at the price of working out four helicity amplitudes instead of a unique amplitude in the case of the SF. The number of helicity amplitudes even increases by increasing the number of spinning particles in the process, such as for instance $e^{+} e^{-} \to \gamma \gamma $ which involves sixteen helicity amplitudes $T_{h}^{\lambda } {}_{h'}^{\lambda '} $.

\noindent

\subsection {The quark dipole magnetic moment}

\noindent

\noindent
By using the Gordon decomposition we divide the dipole magnetic moment expression into two terms: one is the convection current part and the other is the spin part. As an application of our formalism we compute the convection part as it is the only part which involves the generalized spin density. Let us first show that the spin current part of the dipole magnetic moment does not involve the generalized spin density. The spin current part is proportional to

\begin{equation} \label{ZEqnNum319476}
\left. \int \left[\vec{\nabla }_{q} \times (Tr(q_{\nu } \vec{\sigma }^{\nu } \psi (k)\bar{\psi }(k^{'} ))\right] \right|_{\vec{q}=0} \frac{d^{3} k}{(2\pi )^{3} }
\end{equation}
There are two terms resulting from the differentiation with respect to $\vec{q}$. The first one leads to the usual projector $\psi (k,s)\bar{\psi }(k,s)$ after setting $\vec{q}=0$, while the second one is proportional to $\vec{q}$ hence vanishing at $\vec{q}=0$. It then follows that the generalized projector $\rho $ is effectively absent in the expression(\ref{ZEqnNum319476}). The convection current part has the form

\begin{equation} \label{ZEqnNum415208}
\begin{array}{l} {-i\int \left[Tr\vec{\nabla }_{k} \rho \right]_{\begin{array}{l} {\vec{k}'=\vec{k}} \\ {\vec{\zeta }'=\vec{\zeta }} \end{array}} \times \vec{k}\frac{d^{3} k}{(2\pi )^{3} }  } \\ {\, } \end{array}
\end{equation}
In computing the above expression we use the identities $\vec{\nabla }_{k} k_{0} =\frac{\vec{k}}{k_{0} } $ and $\vec{\nabla }_{k} |\vec{k}|=\frac{\vec{k}}{|\vec{k}|} $ to eliminate all differentiations leading to terms proportional to $\vec{k}$ as this leads to $\vec{k}\times \vec{k}=0$ in (\ref{ZEqnNum415208}). In particular any differentiation of the spin variable is eliminated as $\rlap{$/$}s(\rlap{$/$}s')$ is function of the boost parameter $\omega =-\tanh ^{-1} (\frac{|\vec{k}|}{k_{0} } )$($\omega '$) and hence necessarily leads to $\vec{k}$.With these remarks only the variables $\rlap{$/$}k$and$\rlap{$/$}k'$ are differentiated and we get

\begin{equation} \label{ZEqnNum704103}
\begin{array}{l} {\frac{i}{2(k_{0} +m)} \int (\vec{k}\times Tr\left[(\frac{\rlap{$/$}k+m}{2m} )(\frac{1+\gamma _{5} \rlap{$/$}s}{2} )(\gamma _{0} \vec{\gamma })\right])\frac{d^{3} k}{(2\pi )^{3} }  } \\ {=\frac{-i}{8m(k_{0} +m)} \int (\vec{k}\times Tr\gamma _{5} \rlap{$/$}k\rlap{$/$}s\gamma _{0} \vec{\gamma })\frac{d^{3} k}{(2\pi )^{3} }  } \\ {=-\frac{1}{2m(k_{0} +m)} \int |\vec{k}|^{2} \vec{s}_{\bot } (k)\frac{d^{3} k}{(2\pi )^{3} }  } \end{array}
\end{equation}

\noindent Note the natural occurrence of the transverse spin $\vec{s}_{\bot } $ in the calculation of the convection current and hence there is room for the HF. Only the few steps in(\ref{ZEqnNum704103}) are needed in the  computation of  the dipole moment and this has to be compared with the lengthy computation of the same observable using the Dirac spinors\cite{mus6}

\section {Conclusion}

\noindent

\noindent We constructed the SF in its version where the amplitude is expressed as a trace over Dirac indices. Such formalism is appropriate to processes involving transversity knowing that such   processes are now of  great popularity. A similar formalism but where the helicity is the principal entity already exits but it is more appropriate to processes involving states of definite helicity. Although the HF may also deal with processes with transversity too, it does so but only at the price of computing several helicity amplitudes.We also have used the SF to retrieve all formulas of the HF  in an intuitive and clear way. On the other hand we have computed some processes with transverse polarizations, in both formalisms to illustrate the performance of the spin over the helicity in such processes. Finally the dipole magnetic moment as an example is shown to be exclusively  expressed in term of  spin and not of helicity.

\noindent

\appendix*
\section{}
\noindent The idea is to relate the projector $\rho$ directly to its form within the rest frame, where this form is relatively easy to compute. Then to get the desired projector  we perform a Lorentz boost to the moving frame. To do so, let us rewrite the Dirac spinor and it complex conjugate as:

\begin{equation} \label{57)}
\begin{array}{l} {u(k',s{\rm '})=\exp (-\frac{\omega }{2} \gamma _{0} \frac{\vec{\gamma }.\vec{k}'}{|\vec{k}'|} )\left(\begin{array}{c} {\chi _{s{\rm '}} } \\ {0} \end{array}\right)} \\ {\bar{u}(k',s{\rm '})=\left(\begin{array}{cc} {\chi _{s{\rm '}} '^{\dag } ,} & {0} \end{array}\right)\exp (-\frac{\omega }{2} \gamma _{0} \frac{\vec{\gamma }.\vec{k}'}{|\vec{k}'|} )\gamma _{0} } \end{array}
\end{equation}
$\rho$ can now be re-written in terms of its expression in the rest frame

\begin{equation} \label{ZEqnNum728094}
\begin{array}{rcl} {\rho _{s{\rm '}s} } & {=} & {\exp (-\frac{\omega '}{2} \gamma _{0} \frac{\vec{\gamma }.\vec{k}'}{|\vec{k}'|} )\left(\begin{array}{c} {\chi _{s{\rm '}} } \\ {0} \end{array}\right)\left(\begin{array}{cc} {\tilde{\chi }_{s} ^{\dag } ,} & {0} \end{array}\right)\exp (\frac{\omega }{2} \gamma _{0} \frac{\vec{\gamma }.\vec{k}}{|\vec{k}|} )} \\ {} & {=} & {\exp (-\frac{\omega '}{2} \gamma _{0} \frac{\vec{\gamma }.\vec{k}'}{|\vec{k}'|} )\left(\begin{array}{cc} {\chi _{s{\rm '}} \tilde{\chi }_{s} ^{\dag } } & {0} \\ {0} & {0} \end{array}\right)\exp (\frac{\omega }{2} \gamma _{0} \frac{\vec{\gamma }.\vec{k}}{|\vec{k}|} )} \end{array}
\end{equation}
We then work out the expression of the $\chi _{s{\rm '}} \tilde{\chi }_{s} ^{\dag }$ matrix. We give the result without details of the computation

\begin{equation} \label{ZEqnNum636898}
\chi _{s{\rm '}} (\vec{\zeta }')\chi ^{\dag } _{s} (\vec{\zeta })=\frac{{\rm 2{\mathcal N}}_{s's} }{1+s's\vec{\zeta }'.\vec{\zeta }} (\frac{1+s'\vec{\zeta }'.\vec{\sigma }}{2} )(\frac{1+s\vec{\zeta }.\vec{\sigma }}{2} )
\end{equation}
The spin factors ${\rm {\mathcal N}}_{s's} $ will be  given in terms of the helicity factors ${\rm {\mathcal N}}_{\lambda '\lambda } $ which we work seperately

\begin{equation} \label{60)}
\begin{array}{l} {{\rm {\mathcal N}}_{1,1} =\cos \frac{\theta }{2} \cos \frac{\theta '}{2} e^{-i\frac{(\phi -\phi ')}{2} } +\sin \frac{\theta }{2} \sin \frac{\theta '}{2} e^{i\frac{(\phi '-\phi )}{2} } } \\ {{\rm {\mathcal N}}_{-1,1} =-\cos \frac{\theta '}{2} \sin \frac{\theta }{2} e^{-i(\frac{\phi -\phi ')}{2} } +\sin \frac{\theta '}{2} \cos \frac{\theta }{2} e^{i(\frac{\phi -\phi ')}{2} } } \\ {{\rm {\mathcal N}}_{1,-1} =-{\rm {\mathcal N}}^{*} _{-1,1} } \\ {{\rm {\mathcal N}}_{-1,-1} ={\rm {\mathcal N}}^{*} _{1,1} } \end{array}
\end{equation}

\noindent  Inserting(\ref{ZEqnNum636898}) into(\ref{ZEqnNum728094}) we get (in the following we hide for clarity, all occurrence of $s$ and $s'$ and also the factor $\frac{{\rm 2{\mathcal N}}_{s's} }{1+s's\vec{\zeta }'.\vec{\zeta }} $ till the end of the computation)

\noindent

\begin{equation} \label{ZEqnNum593194}
\begin{array}{l} {\rho =\exp (-\frac{\omega '}{2} \gamma _{0} \frac{\vec{\gamma }.\vec{k}'}{|\vec{k}'|} )\left(\begin{array}{cc} {(\frac{1+\vec{\zeta }'.\vec{\sigma }}{2} )(\frac{1+\vec{\zeta }.\vec{\sigma }}{2} )} & {0} \\ {0} & {0} \end{array}\right)\exp (\frac{\omega }{2} \gamma _{0} \frac{\vec{\gamma }.\vec{k}}{|\vec{k}|} )} \\ {=\exp (-\frac{\omega '}{2} \gamma _{0} \frac{\vec{\gamma }.\vec{k}'}{|\vec{k}'|} )\left(\begin{array}{cc} {(\frac{1+\vec{\zeta }'.\vec{\sigma }}{2} )} & {0} \\ {0} & {0} \end{array}\right)\left(\begin{array}{cc} {(\frac{1+\vec{\zeta }.\vec{\sigma }}{2} )} & {0} \\ {0} & {0} \end{array}\right)\exp (\frac{\omega }{2} \gamma _{0} \frac{\vec{\gamma }.\vec{k}}{|\vec{k}|} )} \end{array}
\end{equation}
Now we insert the identity 1(written as the product of four exponentials) between the two spin matrices in(\ref{ZEqnNum593194}) and get

\begin{equation} \label{ZEqnNum159034}
\begin{array}{l} {\rho =\exp (-\frac{\omega '}{2} \gamma _{0} \frac{\vec{\gamma }.\vec{k}'}{|\vec{k}'|} )\left(\begin{array}{cc} {(\frac{1+\vec{\zeta }'.\vec{\sigma }}{2} )} & {0} \\ {0} & {0} \end{array}\right)\exp (\frac{\omega '}{2} \gamma _{0} \frac{\vec{\gamma }.\vec{k}'}{|\vec{k}'|} )} \\ {\exp (-\frac{\omega '}{2} \gamma _{0} \frac{\vec{\gamma }.\vec{k}'}{|\vec{k}'|} )\exp (\frac{\omega }{2} \gamma _{0} \frac{\vec{\gamma }.\vec{k}}{|\vec{k}|} )} \\ {\exp (-\frac{\omega }{2} \gamma _{0} \frac{\vec{\gamma }.\vec{k}}{|\vec{k}|} )\left(\begin{array}{cc} {(\frac{1+\vec{\zeta }.\vec{\sigma }}{2} )} & {0} \\ {0} & {0} \end{array}\right)\exp (\frac{\omega }{2} \gamma _{0} \frac{\vec{\gamma }.\vec{k}}{|\vec{k}|} )} \end{array}
\end{equation}
Now the second line in(\ref{ZEqnNum159034}) is the $\Re$ matrix and the first and the third lines in(\ref{ZEqnNum159034}) are the usual projectors (This is a lengthy but standard computation, so we just give the result)

\begin{equation} \label{ZEqnNum963459}
\begin{array}{l} {\exp (-\frac{\omega }{2} \gamma _{0} \frac{\vec{\gamma }.\vec{k}}{|\vec{k}|} )\left(\begin{array}{cc} {(\frac{1+\vec{\zeta }.\vec{\sigma }}{2} )} & {0} \\ {0} & {0} \end{array}\right)\exp (\frac{\omega }{2} \gamma _{0} \frac{\vec{\gamma }.\vec{k}}{|\vec{k}|} )} \\ {=(\frac{\rlap{$/$}k+m}{2m} )(\frac{1+\gamma ^{5} \rlap{$/$}s}{2} )\, } \end{array}
\end{equation}

\vspace{7mm}

\providecommand{\noopsort}[1]{}\providecommand{\singleletter}[1]{#1}%

\end{document}